# Enhanced quantum coherence in graphene caused by Pd cluster deposition

Yuyuan Qin[1,a)], Junhao Han[1,a)], Guoping Guo[2], Yongping Du[1], Zhaoguo Li[1], You Song[4], Li Pi[3], Xuefeng Wang[5], Xiangang Wan[1], Min Han[1], Fengqi Song[1,b)]

[1] National Laboratory of Solid State Microstructures, Collaborative Innovation Center of Advanced Microstructures, and Department of Physics, Nanjing University, Nanjing 210093, P. R. China

[2] Key Lab of Quantum Information, CAS, University of Science and Technology of China, Hefei, 230026, China

[3] High Magnetic Field Laboratory, Chinese Academy of Sciences, Hefei, 230027, China

[4] Collaborative Innovation Center of Advanced Microstructures, State Key Laboratory of Coordination Chemistry, and School of Chemistry and Chemical Engineering, Nanjing University, Nanjing 210093, P. R. China

[5] School of Electronic Science and Engineering and National Laboratory of Solid State Microstructures, Nanjing University, Nanjing, 210093, China.

[a] Yuyuan Qin, Junhao Han contributed equally

[b] Corresponding Author, songfengqi@nju.edu.cn; Fax +86-25-83595535

Keywords

Graphene; Cluster deposition; Zero-temperature dephasing; weak localization

**Abstract**

We report on the unexpected increase in the dephasing lengths of a graphene sheet caused by the deposition of Pd nanoclusters, as demonstrated by weak localization measurements. The dephasing lengths reached saturated values at low temperatures. Theoretical calculations indicate the p-type charge transfer from the Pd clusters, which contributes more carriers. The saturated values of dephasing lengths often depend on both the carrier concentration and mean free path. Although some impurities are increased as revealed by decreased mobilities, the intense charge transfer leads to the improved saturated values and subsequent improved dephasing lengths.

The surface decoration of graphene offers great opportunities because graphene is a fully open system.[1,2] Functional defects, p/n type doping and additional spin–orbit interactions can be introduced when atoms are absorbed from an external source.[3-12] For example, researchers are even considering inducing topologically nontrivial gaps inside the Dirac cone. It has been suggested that magnetic attachments that break the time reversal symmetry of graphene can induce pseudospin organization, which would result in topologically nontrivial states.[7,13] Adatom adsorptions (In, Tl and Pb) are believed to generate a gap within the Dirac cone, thereby introducing the quantum spin Hall effect in graphene sheets.[11] Ir and W (5d) atom attachments can mediate a quantum anomalous Hall state.[14] An experiment has shown that spin-flip scattering can be introduced by Fluorine.[6] The Au deposits alter spin transport in graphene.[10] These recent developments shed light on how the coherent states can be achieved.

Despite these potential advances, however, an important problem remains that surface adsorption, along with introducing the required functionality, induces additional electronic scattering.[15-17] Such scattering may suppress the coherence of the Dirac fermions and may even disable these desired quantum states. In this Letter, we deposited Pd nanoclusters on graphene sheets and studied the weak localizations (WLs). We found that the dephasing lengths could increase even after the surface was decorated. We propose that decorating graphene with Pd clusters simultaneously induces scattering and introduces charges, leading to unexpected improvements in electronic coherence.

Graphene samples were cleaved from Kish graphite flakes and on a silicon wafer with 300 nm thick silicon wafer .[1] A standard lift-off process was used to fabricate the four-probe electrodes with a back-gate. The beam of Pd clusters was generated in a gas-aggregation cluster source.[18,19] Generally, The coverage is less than 20%. *Defect-enriched* graphene samples were tentatively used with the mobility of less than 1000cm$^2$/Vs. The WL characteristics were analyzed according to the theory used in previous research.[6,20,21] The parameters for the samples used in this study are shown in **Table 1**.

The dephasing length $L_\varphi$ is a critical measurement of electronic interference, governing the fidelity of the electronic phase and quantum entanglement. The WLs in graphene have been intensively studied,[15,16] and WLs were observed in our samples. For instance, the inset in **Figure 1a** shows the typical four-probe measurement configuration used for our graphene samples, which were normally over 5 μm wide and with channels 2μm long. Raman investigations (**Figure 1a**) show that our samples consisted of a single layer of graphene.[17] The magnetoconductance (MC) curve shown in **Figure 1b** provides evidence for the presence of WLs in Sample 1. A normal parabolic magnetoresistance (MR) curve was obtained in the high-field range, but the sharp cusp was dominant near the zero fields, giving a negative MC contribution. The cusp depth decreased as the temperature increased from 2 to 64 K, and the cusp almost disappeared altogether at 150 K, confirming its quantum mechanical origin. A dephasing length of 110 nm was found for this sample at 2 K.

The quantum coherence was improved following the adsorption of Pd nanoclusters. **Figure 2a** shows the typical morphology of Sample 1 after the cluster deposition. The scanning electron microscopy (SEM) image shows that nanoclusters were dispersed on the graphene surface with a very low coverage of less than 10%. X-ray photoemission spectra showed that there were no strong bonds between the Pd and carbon atoms, indicating that the Pd nanoclusters were attached in a way that resembled physical adsorption. The electronic coherence was improved by the Pd deposits as shown in **Figure 2b**. Much sharper WL cusps were observed for the decorated (**Figure 2b**) than for the undecorated (**Figure 1b**) graphene, although they had comparable depths, even at 3 K, which is a higher temperature than that used to produce **Figure 1b**. Distinct noisy signals can be seen in the magenta and black MC curves in **Figure 2b**. The noisy data showed similar patterns at different temperatures, indicating that the observations were from quantum conductance fluctuations.[22,23] **Figure 2c** shows the WL cusps for Sample1 before and after Pd cluster deposition in the same frame, allowing a direct comparison to be made, and confirming the improved coherence caused by Pd deposition. The dephasing length increased from 110 to 195 nm when Pd was deposited. The inset in **Figure 2c** shows the WL cusps for Sample 2 before and after Pd cluster deposition, and it can be seen that the dephasing length increased from 46 to 75 nm when Pd was deposited. **Figure 3a** shows two temperature-dependent dephasing length curves for Sample 1, the lower curve having been obtained before and the upper curve after cluster deposition. The dephasing lengths after deposition were nearly double those before deposition over the whole temperature range. Dephasing lengths are often related to the mean free inelastic electronic scattering length.[22,23] The adsorption of Pd clusters by graphene always introduces additional surface scattering and is expected to result in decreased dephasing lengths. For Sample 1(2), after cluster deposition, the carrier mobility changed from 956 cm$^2$/Vs (300cm$^2$/Vs) to 956 cm$^2$/Vs (200cm$^2$/Vs). It means Pd clusters do not introduce much scattering. But the charge density increased from 0.4×10$^{13}$ cm$^{-2}$ (0.7×10$^{13}$ cm$^{-2}$)to 0.8×10$^{13}$ cm$^{-2}$(1.4×10$^{13}$ cm$^{-2}$), it results the improvement of depahsing lenghths, which is similar to the graphene with variety of charge concentrations by gating and has been well explained by electron−electron interaction.[24]

**Figure 3a** also shows the saturation of the dephasing lengths at low temperatures. The graphene sheet dephasing lengths increased with decreasing temperature, and reached saturated values, both before and after cluster deposition, and this was clearly observed in our samples. The temperature-dependent dephasing lengths can be approximate by the empirical formula[23]

$$\frac{1}{L_\varphi^2(T)} = \frac{1}{L_{\varphi 0}^2} + A_{ee}T + A'_{ee}T^2 \ , \quad (1)$$

in which L$_{\varphi 0}$ is the saturated value, T is the temperature, and A$_{ee}$ and A′$_{ee}$ are fitting parameters. The zero-temperature L$_\varphi$ saturation, which is essentially the zero-temperature dephasing (ZTD) of the Dirac fermions, is significant in graphene.[15,25,26] In previous studies of graphene it was predicted that L$_\varphi$ is infinite at zero temperature, and L$_\varphi$ is believed to be limited by the dimensions of the graphene sheet.[20,25,26] Most recently, some saturation was observed, and this was interpreted to be spin-flip scattering.[6,27] The ZTD also conflicts with traditional e–e interaction theory, which predicts diverging $L_\varphi$ values with decreasing T and an infinite coherence (or dephasing) period at T = 0 K. We examined three fitting parameters for the two curves shown in **Figure 3a**, and all were quite similar, except for the dephasing length saturated values $L_{\varphi 0}$. Therefore, improving the quantum coherence after Pd cluster deposition should be investigated in terms of the ZTD.

It would be appropriate to determine whether there could be unexpected fluctuations in the dephasing length at lower temperatures, so the WL measurements were carried out down to 50 mK. The inset of **Figure 3b** shows the WL curves for Sample 3 from 64 K to 50 mK. The WL dips became more obvious as the temperature decreased. However, all the WL curves seem coincided at 4 K and below. Fitting the WL curve gave the temperature dependence of the dephasing length, which saturated at 90 nm. The dephasing length saturated at 700 nm for a sheet (sample 4) with a higher mobility of 2400 $cm^2$/Vs. All the presented evidence indicates the universal existence of the ZTD.

As described above, the conductivities of most of our samples increased when clusters were deposited. Tuning the properties by putting atom or cluster on graphene attracted a lot of research attention. For example, it has been found that putting Co atom onto graphene will change its electronic property dramatically, and the system shows Kondo like behavior. Thus, it is important to clarify the effect of adding Pd atoms onto graphene, which is addressed by density functional calculation. The

Density Functional Theory (DFT) calculations were carried out to calculate the electronic structures of graphene and Pd-cluster-decorated graphene using the plane wave pseudopotential method, which was applied using the Vienna *ab initio* simulation package. The graphene superlattice was made up of 6 × 6 graphene unit cells with Pd clusters absorbed. The structures were optimized by relaxing the positions of the ions until the Hellmann–Feynman forces were less than 0.05 eV/Å.

The clusters used in the calculations contained 1, 3, 5, 7, and 9 Pd atoms. A typical model was the cluster $Pd_3$ supported on graphene sheet as shown in **Figure 4**. The calculation revealed the Dirac cone remained almost intact while the Fermi level moved downward because of the intense charge transfer. Calculations for all the clusters confirmed that charge transfer occurred, predicting an increase in the conductivity and carrier density when the clusters were deposited. No spin moments greater than 0.001 $\mu_B$/atom were seen in the C electronic states. The charge transfer was reasonable because of the difference in the work function between Pd and graphene. Note that the clusters in the experiments contained more than 10000 Pd atoms, so each cluster would have magnetism parameters more similar to the bulk metal than to clusters of only a few Pd atoms. Pd metal is nonmagnetic, so the exclusion of spin-flip scattering, and even non-spin-flip scattering after Pd decoration, is reasonable as described above. The calculation demonstrated the intense p-type charge transfer and further excluded spin transfer.

The adsorption of Pd clusters introduced both additional scattering and charge transfer to the graphene sheet. Our samples were normally slightly p-doped because of the gold contacts used. The deposition of Pd clusters therefore further increased the conductivity of the graphene sheets. The scattering suppressed the quantum coherence and the additional charge suppressed the ZTD, increasing the upper boundary of the coherent period of the Dirac fermions. The final coherence improvement was a result of competition between scattering suppression and ZTD. The coherence improvement accompanied with the charge transfer may be useful in the solar cell systems based on graphene composites. One may add nanocluster-decorated graphene composites into the solar cells[28-30]. Besides the charge transfer, the cluster decoration also improves the electronic coherence, which has been believed to the time scale of the energy transfer in an optoelectronic procedure[31,32]. The cluster decoration therefore may critically improve the quantum efficiency of the solar electronic devices.

In summary, the graphene sheet's dephasing lengths increased because of the deposition of Pd nanoclusters, as shown by the weak localizations. The temperature-dependent and zero-temperature saturated dephasing period data were interpreted using a simple relationship considering the intrinsic e–e interactions. The increase in the dephasing length was attributed to the increase charge concentration. The cluster decoration of graphene not only induced additional scattering but also caused more intense charge transfer. The approach described here allows the Dirac carrier coherence to be retained while functional impurities are introduced into the graphene. It is important to use nanoclusters rather than individual atoms as the "decorations" because this leads to less scattering but still allows the same amount of control over electronic interference.

**Acknowledgments**

We thank the National Key Projects for Basic Research of China (Grant Nos: 2013CB922103, 2011CB922103, 2014CB921103), the National Natural Science Foundation of China (Grant Nos: 91421109, 11023002, 11134005, 61176088, 11222438 and 11274003), the PAPD project, The Natural Science Foundation of Jiangsu Province (Grant BK20130054) and the Fundamental Research Funds for the Central Universities for their financial support of this work. We would like to acknowledge the helpful assistance of the Nanofabrication and Characterization Center at the Physics College of Nanjing University, Prof. Wei Ren at Shanghai University, Prof. Zengfeng Di at Shanghai Institute and Microsystems, Prof. Y. Q. Li at institute of Physics, and Dr. Ziyou Li at the University of Birmingham (U.K.).

**Tables**.

| Sample | R (Ω) | Mobility (cm$^2$/Vs) | Le (nm) | n (10$^{13}$/cm$^2$) | Lφsat (nm) |
|---|---|---|---|---|---|
| 1 | 812/426 | 956/956 | 22.8/31.4 | 0.4/0.8 | 110/195 |
| 2 | 737/554 | 300/200 | 9.3/8.7 | 0.7/1.4 | 46/75 |
| 3 | 3300 | N.A. | N.A. | N.A. | 90 |
| 4 | 81 | 2400 | 57 | 0.3 | 700 |
| 5 | 216 | 725 | 21 | 0.6 | 113 |

**Table 1** Parameters for the five samples described in the manuscript. Where there are two values in one box these are the values before (left) and after (right) Pd deposition. N.A. means that the value is not available because there was no parabolic MR even at a higher field.

**Figure Caption**

**Figure 1. Sample quality and the magnetoresistance (MR) measurements of a graphene sheet before Pd clusters were deposited (sample 1).** (a) The Raman spectrum of a typical sample. The 2D peak and the 2D/G ratio indicate that the sample was a single layer graphene sheet. The inset shows the optical micrograph of the sample and the four-probe measurement configuration, with a scale bar of 4 μm. The dashed lines show the edges of the graphene sheet. (b) The relative magnetic conductance, ΔG(B) (= G(B)−G(0)), plotted against the perpendicular magnetic field B, at various temperatures. Zero-field cusps, manifesting weak localizations (WLs), can be seen. The weak localizations decayed with increasing temperature.

**Figure 2. The surface morphology and magnetoresistance (MR) response of the graphene sheet after Pd clusters had been deposited (sample 1).** (a) A scanning electron microscopy image of the graphene sheet after Pd clusters had been deposited. The inset shows an atomic force microscopy image of the gold electrode. The clusters appear to be granular and discrete. (b) The relative magnetic conductance ΔG(B), plotted against the perpendicular field B, at various temperatures. (c) Focusing on the weak localization (WL) cusp at low temperatures. The two curves are the MR curves measured at 2 K before deposition (in blue) and at 3 K after deposition (in red). The cusp became much sharper after the Pd clusters had been absorbed, clearly indicating an improvement in electronic coherence. A similar improvement can be seen in the

inset, which shows the MR curves of Sample 2 before and after Pd clusters were deposited (measured at 3 K).

**Figure 3. The temperature dependence of electronic dephasing: zero-temperature saturation.** (a) The temperature-dependent dephasing lengths $L_\varphi$, derived from the weak localization (WL) formula fitting. $L_\varphi$ became saturated at low temperatures, for which the fittings, shown as smooth lines, gave the expected saturated values at 0 K. $L_\varphi$ systematically increased because of the cluster decoration. (b) The WL measurement of Sample 3 extended to 50 mK, at which temperature the ZTD was still found. The inset shows the measured MC curves.

**Figure 4. Simulation of the electronic band** (a) The supercell employed in the simulation, where a cluster with 3 Pd atoms is attached on a 6x6 unit cell of graphene. (b) The band structure obtained by the simulation, where 0 means the Fermi energy. One may find the shift of the Fermi energy towards p-type doping.

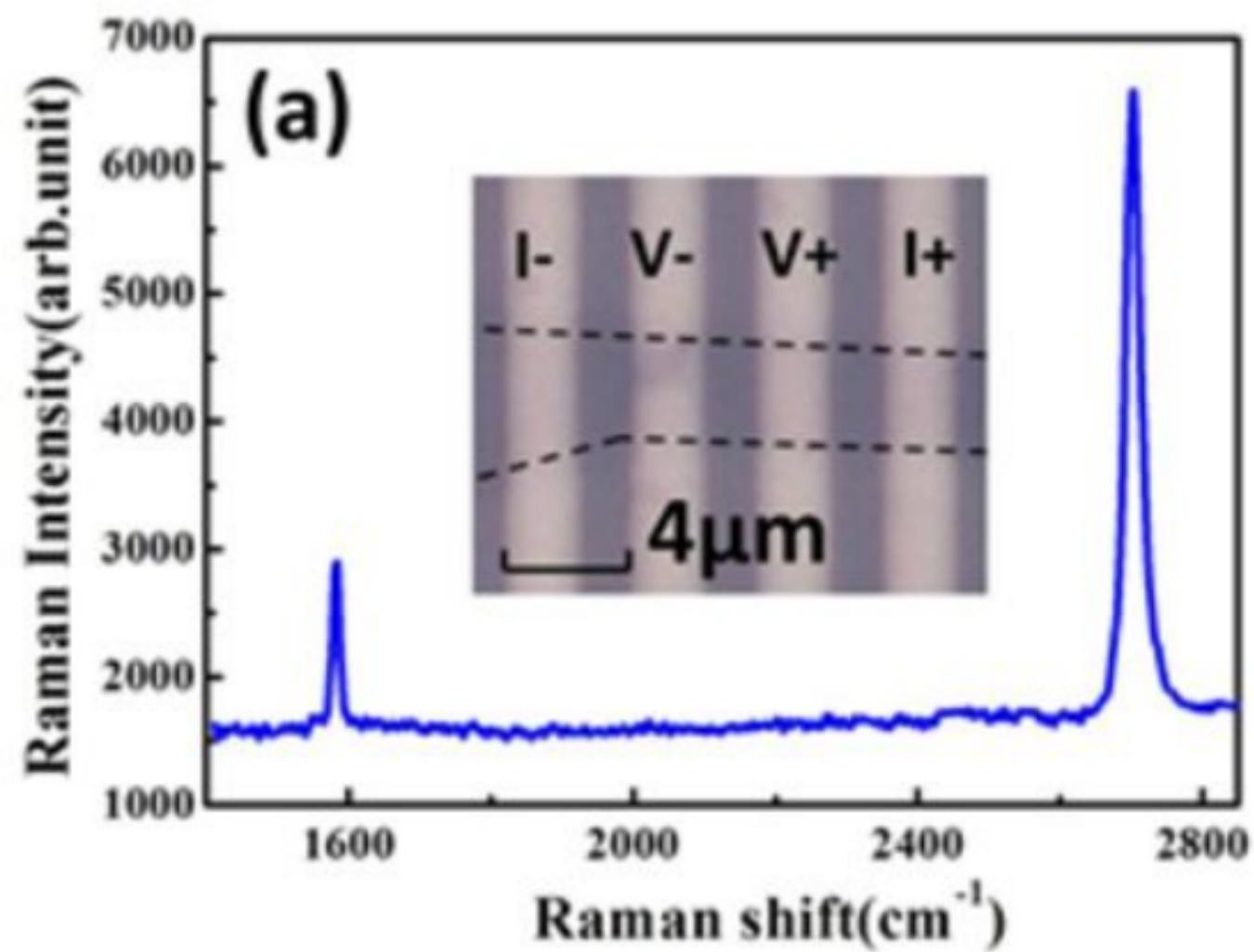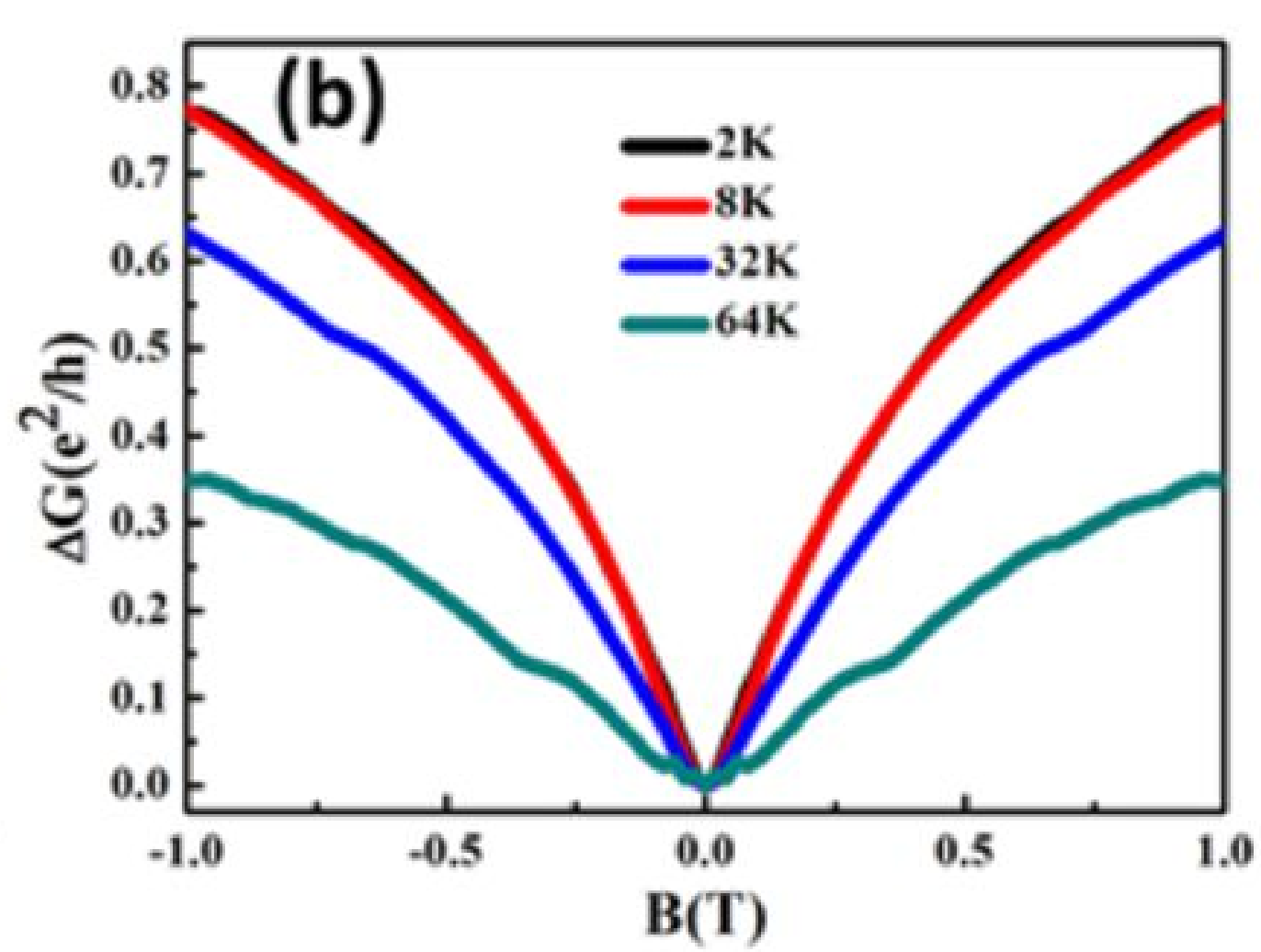

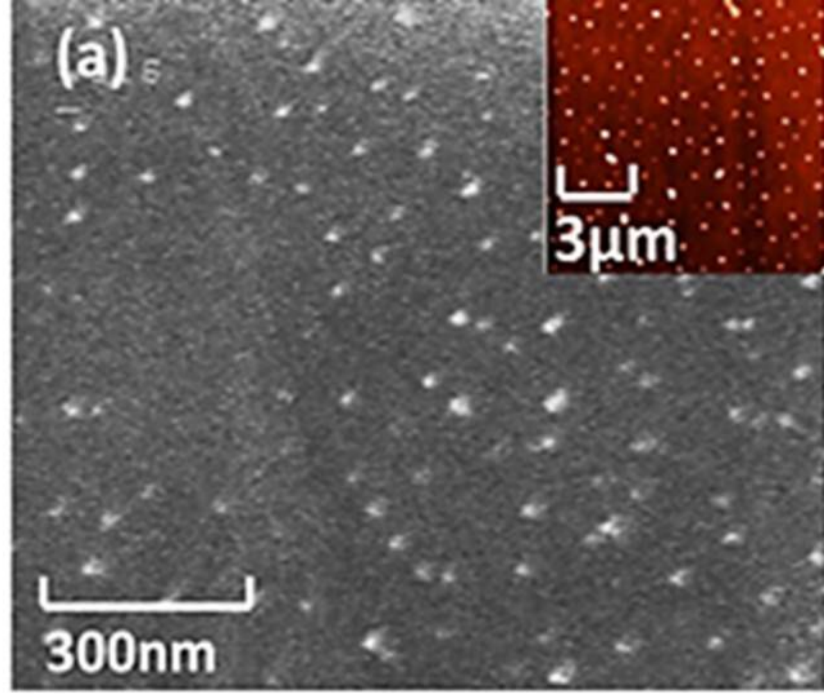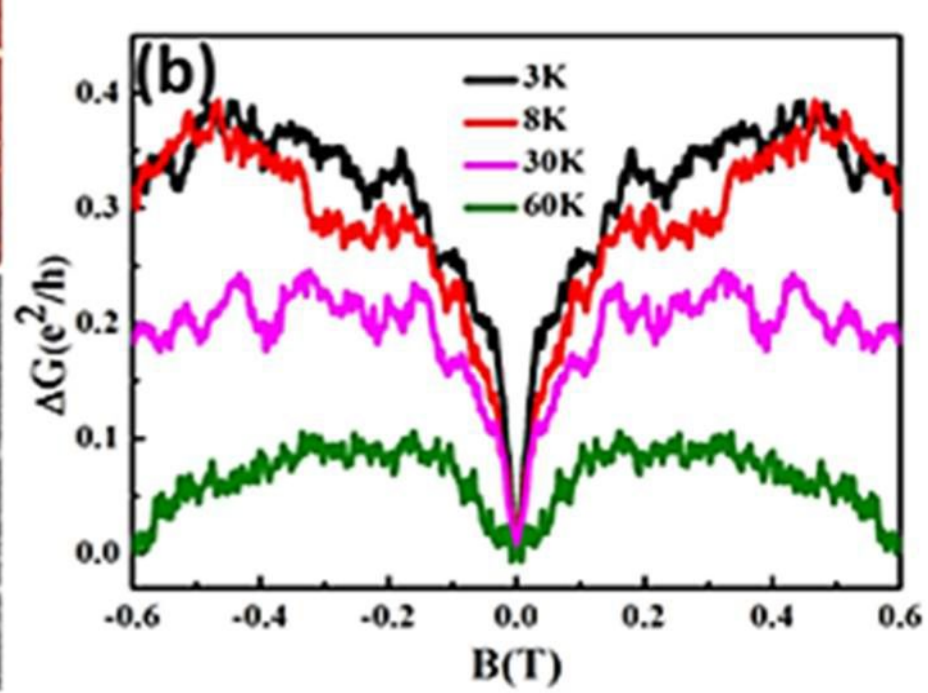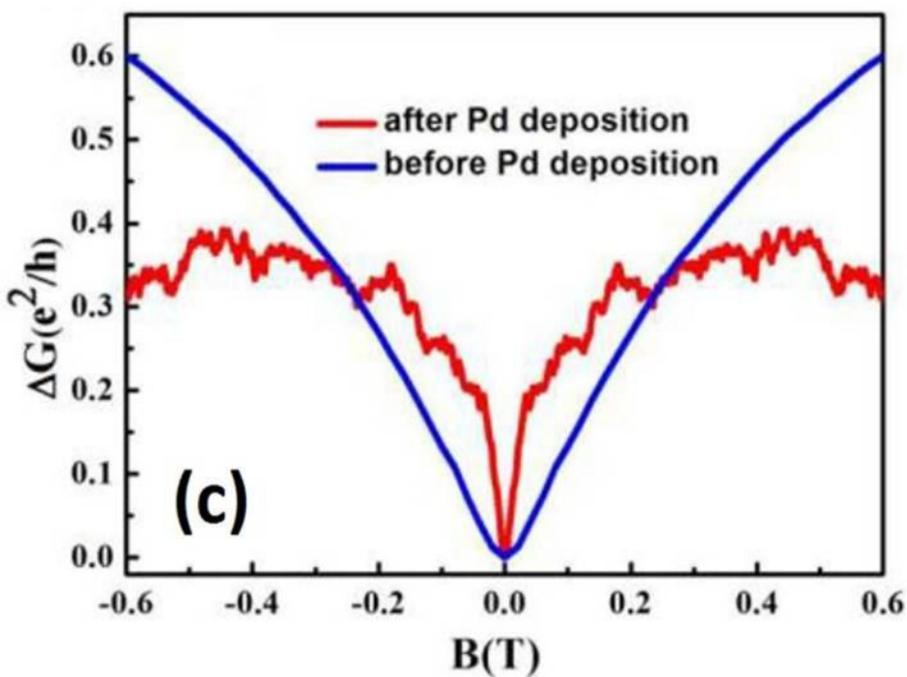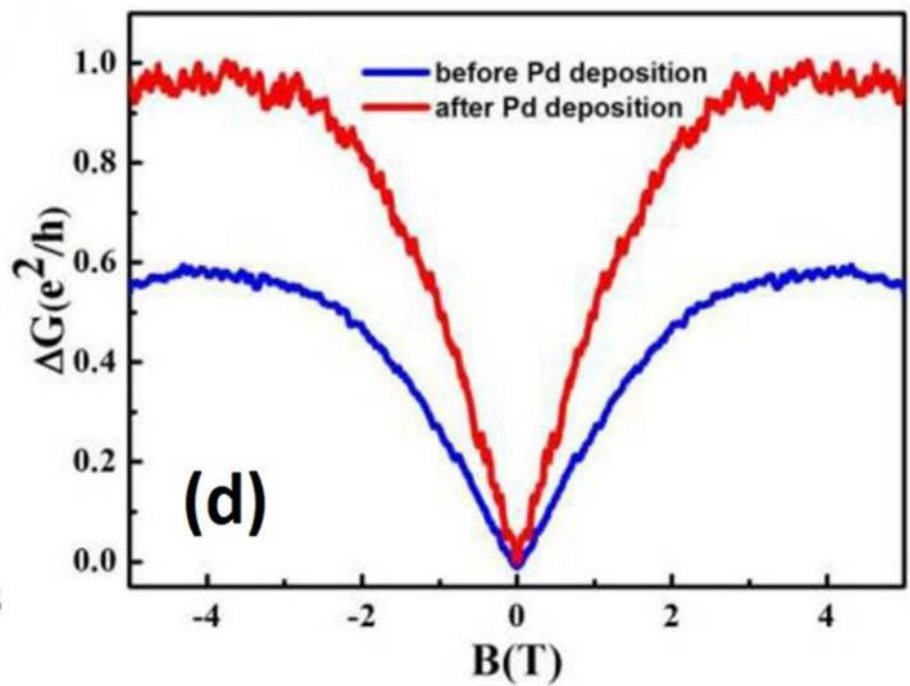

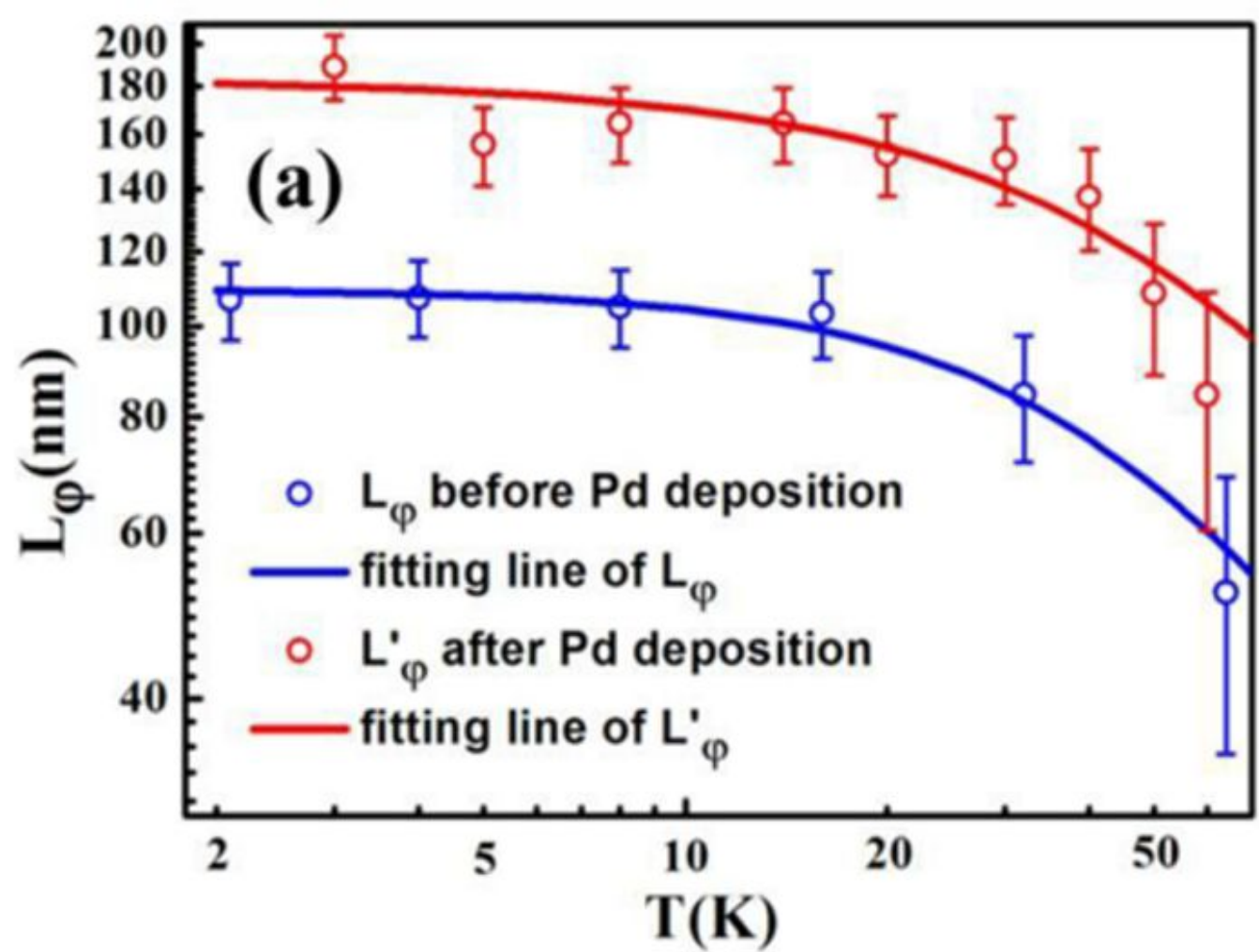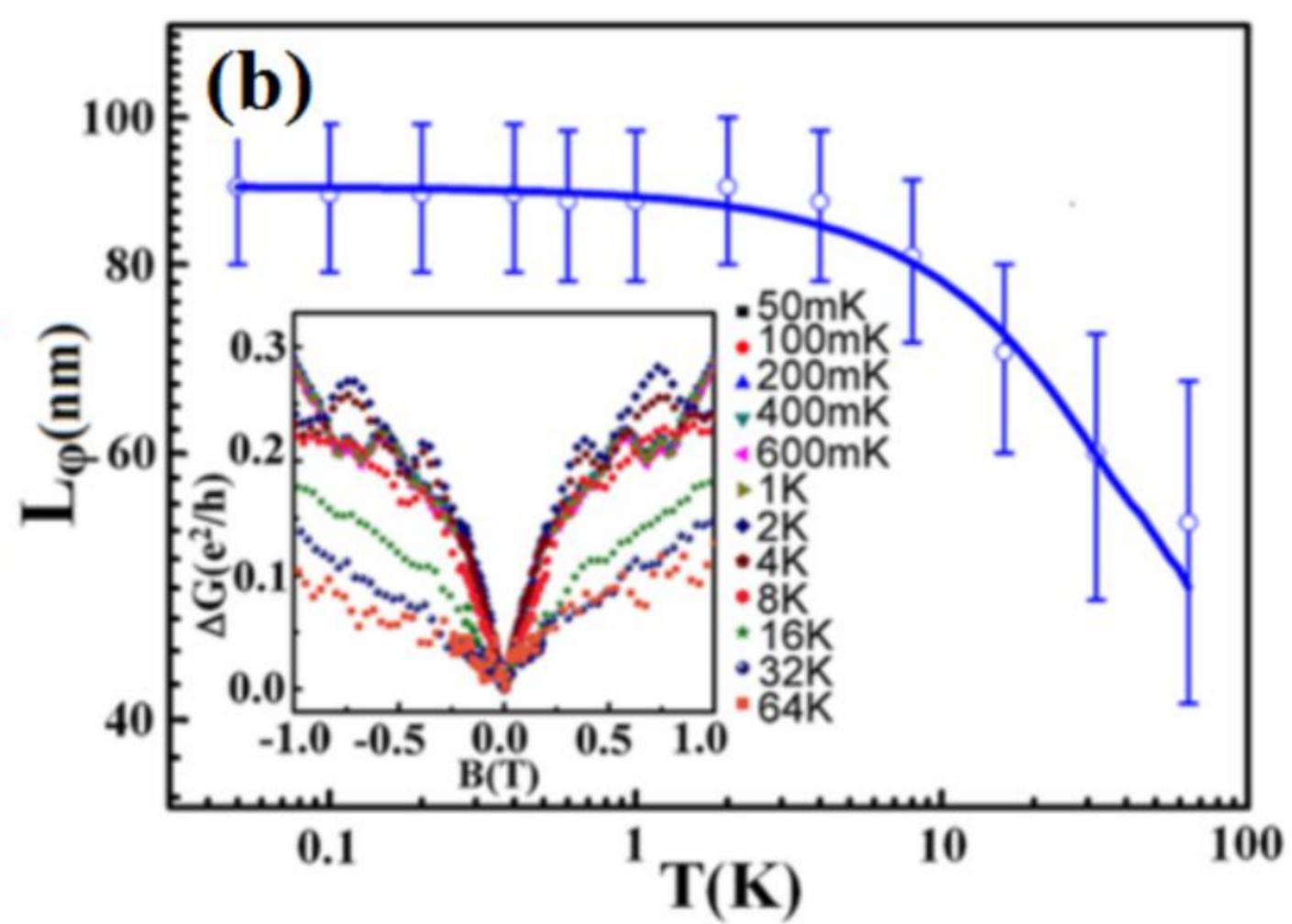

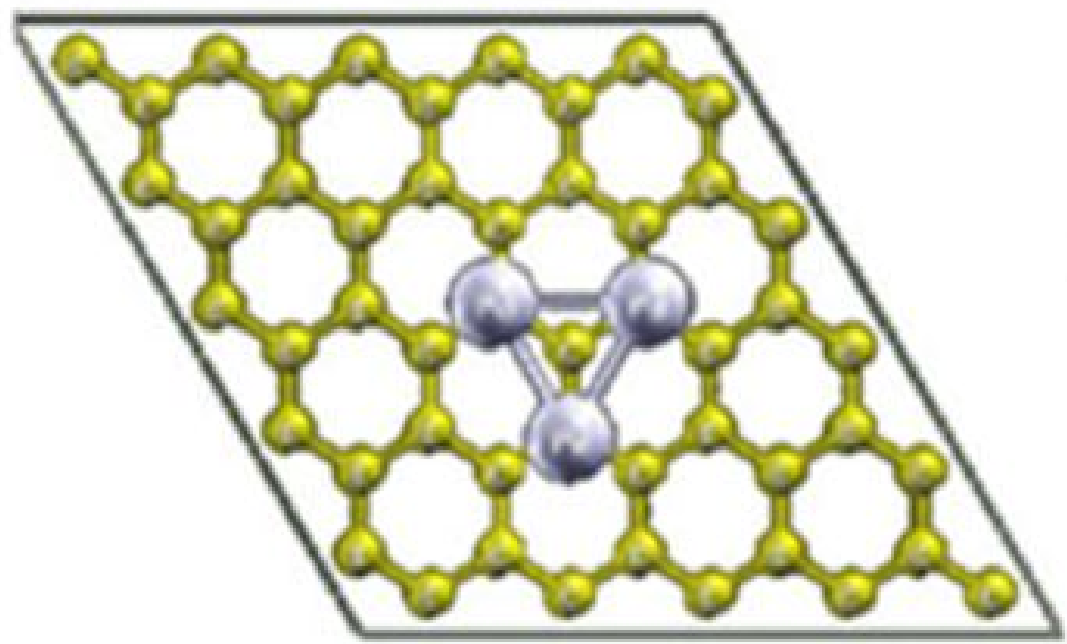 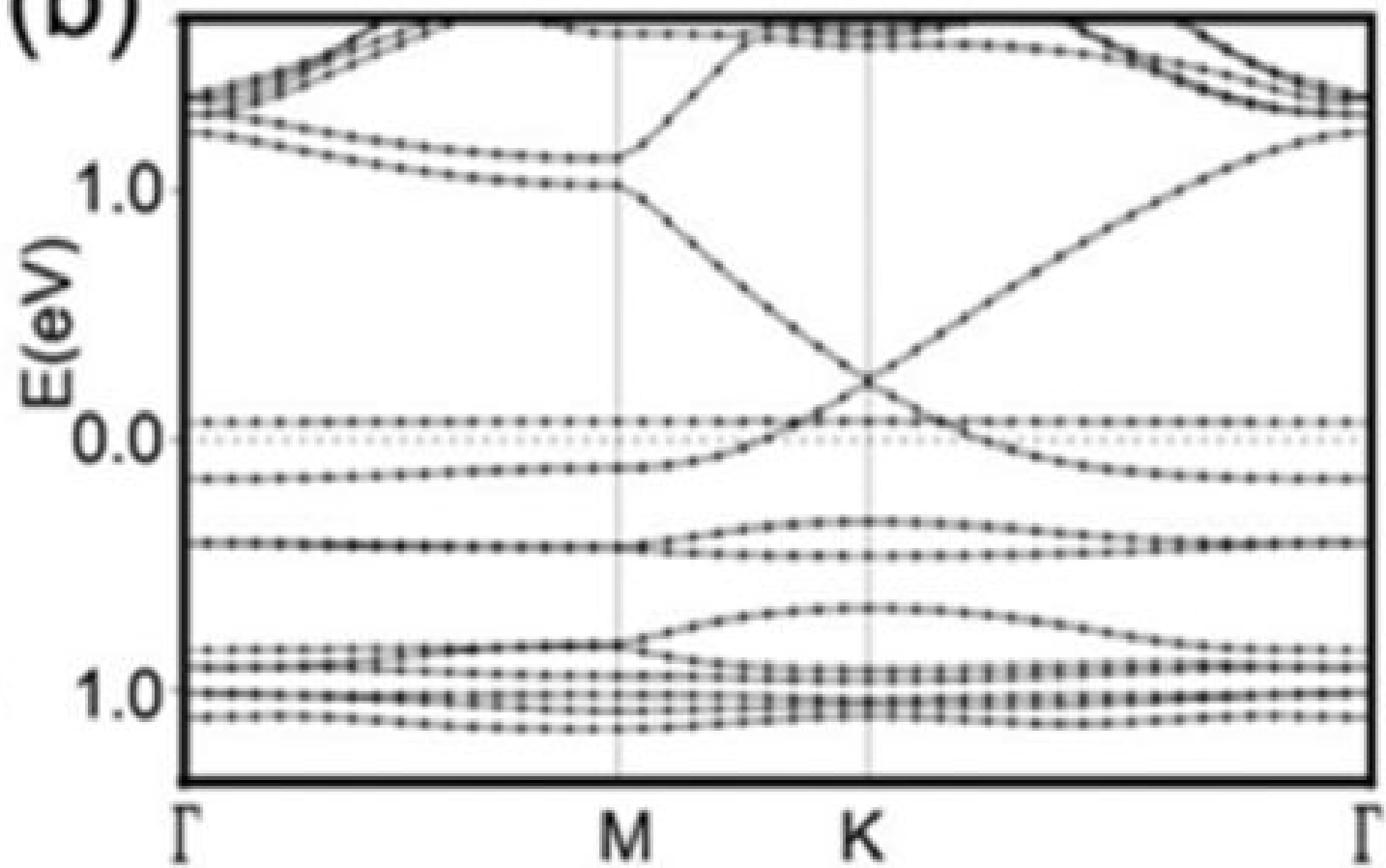